\documentclass[12pt]{article}
\usepackage{epsf}

\newlength{\dinwidth}
\newlength{\dinmargin}
\def\lapproxeq{\lower .7ex\hbox{$\;\stackrel{\textstyle <}{\sim}\;$}}
\def\gapproxeq{\lower .7ex\hbox{$\;\stackrel{\textstyle >}{\sim}\;$}}

\def\be{\begin{equation}}
\def\ee{\end{equation}}
\def\bea{\begin{eqnarray}}
\def\eea{\end{eqnarray}}

\def\lesim{ \;\raisebox{-.7ex}{$\stackrel{\textstyle <}{\sim}$}\; }

\def\qbar{{\bar q}}
\def\ycut{y_{\rm cut}}

\def\GeV{{\rm GeV}}

\def\ra{ \rightarrow }
\def\bb{{b\bar{b}}}

\def\gaga{\gamma\gamma}

\begin{document}
\titlepage

\begin{flushright}
IPPP/06/45\\
DCPT/06/90\\
11 July 2006 \\
\end{flushright}

\vspace*{4cm}

\begin{center}
{\Large \bf On radiative QCD backgrounds to exclusive \\[2mm]
$H\to b \bar b$ production at the LHC and a photon collider}

\vspace*{1cm} \textsc{V.A.~Khoze$^{a,b}$, M.G. Ryskin$^{a,b}$ and W.J. Stirling$^{a,c}$} \\

\vspace*{0.5cm} $^a$ Department of Physics and Institute for
Particle Physics Phenomenology, \\
University of Durham, DH1 3LE, UK \\[0.5ex]
$^b$ Petersburg Nuclear Physics Institute, Gatchina,
St.~Petersburg, 188300, Russia \\[0.5ex]
$^c$ Department of Mathematical Sciences, 
University of Durham, DH1 3LE, UK \\%
\end{center}

\vspace*{1cm}

\begin{abstract}

Central exclusive Higgs boson production, $pp\to p \oplus H \oplus p$, at the LHC
and s-channel resonant Higgs production in the photon-collider option of the ILC
can provide a very important contribution to the comprehensive study of the Higgs sector.
Especially attractive is the  $b \bar b$ Higgs decay mode, which for certain MSSM scenarios may 
become {\it the} discovery channel in exclusive Higgs production at the LHC and the Photon Collider (PC).
Strongly suppressed and controllable backgrounds is an obvious requirement for the success
of these exclusive measurements. 
One of the main sources of background comes from additional gluon radiation
which leads to a three-jet  $b \bar b g$ final state.
We perform an explicit calculation of the subprocesses  $gg\to q\bar q g$ and $\gamma\gamma\to q\bar q g$,
where the incoming particles are required to be in a $J_z=0$ state and the two gluons
form a colour singlet, and investigate the salient properties of these potentially important background processes.

\end{abstract}

\newpage

% ===========================================================================
\section{Introduction}

The identification of the Higgs boson(s) is one of the main goals of the LHC.
Once the Higgs boson is discovered, it will be of primary importance
to determine its spin and parity, and to measure precisely
the mass, width and couplings. A comprehensive study of the whole Higgs sector,
including precision  mass and coupling measurements, spin and CP properties, will be the next stage.
The conventional strategy to achieve this ambitious programme
 requires an intensive interplay between the LHC and the ILC (high-energy linear $e^+e^-$ collider) \cite{Georg} .
The ILC would enable a comprehensive phenomenological
profile of the Higgs sector to be obtained, see for example \cite{ghk}. In particular, a unique
possibility to produce neutral Higgs bosons exclusively as $s$-channel
resonances is offered by the $\gamma\gamma$ Compton Collider option
of the ILC, see for example \cite{telnov,bbc,km,pn,godbole} and references therein.  
Whilst awaiting the arrival of the ILC, there has been growing interest in recent years
in the possibility to complement the standard LHC physics menu by 
adding forward proton taggers to the CMS and ATLAS experiments (see for example \cite {ar,KMRProsp,DKMOR,
cox1,JE,LOI,bh75,CR,KPR,KKMRext} and references therein).

\begin{figure}[h]
\begin{center}
\vspace{1cm}
\centerline{\epsfxsize=0.4\textwidth\epsfbox{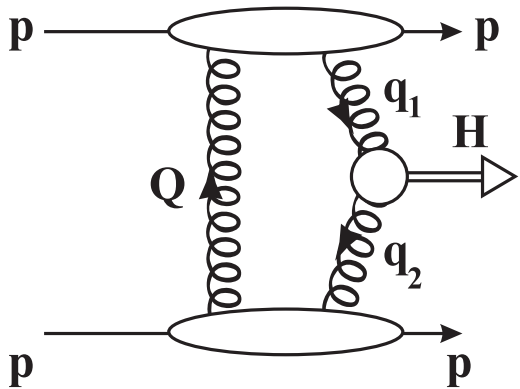}}
\caption{Schematic diagram for central exclusive Higgs production at the LHC,
$pp \to p+H+p$.}  
\label{fig:H}
\end{center}
\end{figure}

While experimentally challenging, this would provide an exceptionally clean environment to search for, and to
identify the nature of, the new objects at the LHC.
One of the key theoretical motivations behind these recent proposals is the study of 
so-called `central exclusive' Higgs boson production, $pp\to p \oplus H \oplus p$.
The $\oplus$ signs are used to denote the presence of large rapidity gaps; here
we will simply describe such processes as `central exclusive', with
`double-diffractive' production being implied. In these exclusive processes there is no
hadronic activity between the outgoing protons and the decay products of the central (Higgs) system.
The predictions for exclusive production are obtained by calculating the diagram of Fig.~\ref{fig:H}
using perturbative QCD \cite{KMR,KMRProsp,jeff}. In addition, we have to calculate and include the probability
that the rapidity gaps are not populated by secondary hadrons from the underlying event \cite{KMRsoft,maor}.

There are three major reasons why central exclusive production is so attractive
for Higgs studies.
First, if the outgoing protons remain intact and scatter through small angles then, to a very good approximation,
the primary active di-gluon system obeys a $J_z=0$, CP-even selection rule
\cite {Liverpool,KMRmm}. Here $J_z$ is the projection of the total angular momentum
along the proton beam axis. This selection rule readily permits a clean determination 
of the quantum numbers of the observed Higgs resonance which
 will be dominantly produced in a scalar state.
Secondly, because the process is exclusive, the energy loss of the outgoing protons is directly
related to the mass of the central system, allowing a potentially excellent mass resolution, irrespective 
of the decay mode of the produced particle.\footnote{Recent studies suggest
\cite{LOI} that the missing mass resolution 
$\sigma$ will be of order $1\%$ for a $140$~GeV Higgs,
 assuming both protons are detected at 420m from the interaction point \cite{RO,cox1}.}
And, thirdly, a signal-to-background ratio of order 1
(or even better) is achievable. As discussed in \cite {khrstw}, central exclusive
production would enable a unique signature for the MSSM 
Higgs sector, in particular allowing the direct measurement of the $Hbb$ Yukawa 
coupling. Moreover, in some MSSM scenarios this mechanism
provides an opportunity for lineshape analysing \cite{KKMRext,JE}, and offers a way
for direct observation of a CP-violating signal in the Higgs sector \cite{KMRCP,JE}.

The analysis in \cite{KMR,DKMOR,KKMRext} was focused primarily
 on light SM and MSSM Higgs production, with the Higgs 
decaying to 2 $b-$jets. The potentially copious $b-$jet (QCD) background is controlled by
a combination of the $J_z=0$
selection rule \cite{Liverpool,KMRmm}, which 
strongly suppresses  leading-order  $b \bar b$ production, colour and spin factors and the 
mass resolution from the forward proton 
detectors. 
It is the possibility to observe directly the dominant $b \bar b$ decay mode 
of the SM Higgs with $M_H\lesim 140$~GeV that first attracted attention to exclusive production at the LHC.
It was subsequently realised that certain 
regions of MSSM parameter space can be especially `proton tagging friendly'.
For example, at large  $\tan\beta$ and $M_H\lesim 250$~GeV
%$140 gtrsim\M_H\lesim 250 GeV$
the situation becomes exceptionally favourable, with predicted Higgs signal-to-background ratios in excess 
of 20 \cite{KKMRext,khrstw}. In this particular case the tagged proton mode may well
be {\it the} discovery channel.
Though from an experimental perspective the  $b \bar b$ channel
is more challenging
than  the $WW$ decay mode (see \cite{cox2,krs1,ma}), its
unique advantages definitely merits a detailed analysis in realistic experimental conditions at the LHC.  

 At the same time, the PC is especially best suited to 
the precise
measurement of the $\Gamma (H\to\gamma\gamma)$ width. Moreover, 
for certain regions  of the MSSM parameter space,
for example at the so-called `LHC wedge', 
the PC has  a discovery potential
for the heavy pseudoscalar and scalar bosons, $A$ and $H$,
see for example \cite{pn}.
It is instructive to recall that in the case of 
  $\gamma\gamma\to H \to b \bar b$ production
the potentially copious continuum $b-$jet background can be controlled
by using polarised photon beams in the  $J_z=0$ initial-state 
(see for example \cite{GH,bbc}), the same configuration
of incoming particle polarisations that `automatically' appears in the 
case of the pQCD box diagram of  Fig.~\ref{fig:H} for forward going protons at the LHC.
The reason is that the Higgs signal is produced by photons (gluons) 
in a $J_z=0$ state whereas the LO backgrounds are primarily initiated by 
the initial states with  $|J_z|=2$, the $J_z=0$ contribution being suppressed for
large angles by a factor $m_b^2/s$, see for example \cite{ispir, GH}. 
As discussed in \cite{VAK1,BKSO} for the $\gamma\gamma$
case, the physical origin of this suppression is
related to the symmetry properties of the Born helicity amplitudes 
$M_{\lambda_1, 
\lambda_2}^{\lambda_q,\lambda_{\bar q}}$ describing the binary background process
\be
\gamma (\lambda_1, k_1) \: + \: \gamma (\lambda_2, k_2) \;
\rightarrow \; q 
(\lambda_q, p) \: + \: \overline{q} (\lambda_{\bar q}, \overline{p}) \/ .
\label{eq:a1}
\ee
Here $\lambda_i$ labels the helicities of the incoming photons, and
$\lambda_q$ and $\lambda_{\bar q}$ are the (doubled) helicities of the
produced quark
and antiquark.  The $k$'s and $p$'s denote the particle
four-momenta, with $s=(k_1+k_2)^2$.

Specifically, for a $J_z = 0$ initial state
$(\lambda_1 = \lambda_2)$ 
the Born quark helicity conserving (QHC) amplitude
with $\lambda_{\bar q} = -\lambda_q$ vanishes,

\be
 M_{\lambda, \lambda}^{\lambda_q, -\lambda_q} \; = \; 0,
\label{eq:a2}
\ee
see also \cite{fkm}.
For the quark helicity non-conserving (QHNC) amplitude for
large angle 
production we have
\be
M_{\lambda, \lambda}^{\lambda_q, \lambda_q} \; \sim \; {\cal O} \left (
\frac{m_q}{\sqrt{s}} 
\right ) \: M_{\lambda, - \lambda}^{\lambda_q, -\lambda_q},
\label{eq:a3}
\ee
where the amplitude on the right-hand-side displays the dominant
helicity configuration of the background process.  
The above-mentioned $m_b^2$ suppression of the $\gamma\gamma (J_z=0)\to b\bar b$
Born cross section is a consequence of Eqs.\ (\ref{eq:a2}) and
(\ref{eq:a3}). 
The same is valid for the leading order amplitude of the $gg^{PP}\to b \bar b$
subprocess, where the notation $gg^{PP}$ indicates that each active gluon
in Fig.~\ref{fig:H} comes from a colour-singlet $t-$channel (Pomeron) exchange
and that the colour singlet di-gluon subprocess obeys the $J_z=0$, parity-even selection
rule.\footnote{It is worth noting that in the massless limit
Eq.~(\ref{eq:a2}) holds for any colour state of initial gluons.
This is a consequence of the general property that the non-zero massless tree-level
amplitudes should contain at least two positive or two negative helicity
states, see for example \cite{mhv1}. It is an example of the more general Maximally Helicity Violating 
amplitude (MHV) rule, reviewed for example in \cite{MP}.}

The $m_b^2/s$ suppression is especially critical in controlling the $b \bar b$ background.
However, as was pointed out in \cite{BKSO}, the
suppression of the $J_z=0$ background cross section is removed by the
presence of an additional gluon in the final state. The radiative three-jet
processes can then mimic the two-jet events in the quasi-collinear
configurations when the gluon is radiated close to the $b-$quark directions\footnote{In the PC case there also a sizeable radiative background coming from
$c \bar c g$ production.}
or (in the $gg^{PP}$ case) the extra gluon goes unobserved in the direction
of a forward proton.
First evaluations
of the NLO QCD radiative backgrounds at the PC were performed in \cite{BKSO,fkm}
(for further development see \cite{pn,srj} 
and references therein).
This  background contribution appears to strongly exceed the LO expectation and
results in  different shapes for various distributions. 
The background situation for the central exclusive $H\to b \bar b$ production at the LHC
is much more complicated and requires a detailed combined study of various effects,
see \cite{DKMOR,bh75}. The analysis of some of these phenomena
is still incomplete and require further detailed theoretical 
efforts, see, in particular, Sections~2.3 and 3 below.

An important ingredient to this complex study which has not been completed so far 
is the availability\footnote{Although a number of automated packages are available for tree-level scattering amplitudes for arbitrary final states, it is very difficult to extract from these the projection onto a specific spin (e.g. $J_z=0$) and colour (e.g. colour singlet) initial state.}   of the analytical expression for the 
matrix elements of the NLO process $gg^{PP}\to b \bar b g$ which are needed
 to perform explicit calculations of the radiative background in the presence of 
 realistic experimental cuts and selections.
It is one of the main aims of this paper to derive the analytical expressions
for the radiative cross sections,
which then can be convoluted with existing Monte Carlo codes \cite{exhume,bhmps}
for the calculation of  central exclusive processes.
Note that, technically, the calculations of both $gg^{PP}$ and $\gamma\gamma$
induced colour singlet processes are quite similar, the latter providing just a subset
of the diagrams for the former. Therefore, it is convenient to discuss the two
radiative processes simultaneously, illuminating their similarities and
differences. These are discussed in Sections 3 and 4.

Our phenomenological discussion will be focused on central
exclusive $b \bar b$ production at the LHC. However it is also worth noting that the CDF Collaboration
at the Fermilab Tevatron is currently performing an experimental study of exclusive diffractive 
$b \bar b$ events \cite {CDFb}, and the results of this paper could prove useful for 
the analysis of these measurements.

\section{On the backgrounds to the $p+(H\ra \bb)+p$ signal}

\subsection{Classification of the backgrounds to exclusive Higgs production}

From the theoretical point of view, it is convenient to consider
separately the QHC and QHNC amplitudes. These amplitudes do not interfere, and
their contributions can be treated independently. This
could be especially convenient at the stage when the parton shower algorithm is applied,
since double counting can be avoided.  

There are two main sources of    $gg^{PP}\ra\bb$  background process:
\begin{itemize}
\item[(i)] the LO ${\cal O}(\alpha_S^2)$ QHNC amplitude squared,
\item[(ii)] the NNLO ${\cal O}(\alpha_S^4)$ QHC contribution which comes from the
one-loop box diagrams.
\end{itemize}

As already mentioned, there is also the possibility of a NLO  ${\cal O}(\alpha_S^3)$ $gg^{PP}\to b \bar b g$
background contribution, because large-angle, hard-gluon radiation does not
obey the selection rules. Of course, the extra gluon may be observed
experimentally in the central detector, and such background events eliminated. However,
there are important exceptions which we discuss below. 

In the case of the NLO $gg^{PP}\to b \bar b g$ process the dominant
contribution comes from the QHC amplitude, since the QHNC piece is additionally
mass-suppressed. Here we consider two types of radiative background process that can mimic the $H \to b \bar b$ central exclusive signal.
 
\begin{itemize}

\item[(a)] The extra gluon may go unobserved in the direction of
one of the forward protons. This background may be reduced by
requiring the approximate equality $M_{\rm missing} = M_{\bb}$.
But the degree of this reduction will depend on the mass resolution
in the proton detector and jet energy resolution in the central detector. Since the mass (jet energy) resolution
$\Delta M_{\bb}$ in the central detector is expected to be much worse than the missing mass resolution, $\Delta M_{\rm missing} \ll \Delta M_{\bb}$, the background will be limited in practice by the $\Delta M_{\bb}$ value.
\item[(b)]The remaining danger is
large-angle hard gluon emission which is collinear with either the
$b$ or $\bar{b}$ jet, and, therefore, unobservable.
% The extra gluon is collinear with either the $b$ or
%$\bar{b}$ jet.
As discussed in \cite{DKMOR,BKSO}, for $J_z=0$ this is
suppressed for {\it soft} gluon radiation. Although there is a certain suppression
of collinear radiation as well, this issue requires further detailed analysis, see below.

According to the study in \cite{DKMOR}, if the cone
angle needed to separate the $g$ jet from the $b$ (or $\bar{b}$)
jet is $\Delta R \sim 0.5$ then the expected background from
unresolved three jet events leads to $B/S \simeq 0.2$.
The calculations presented in Section 3 below will allow a more precise evaluation of this ratio.

It is worth noting that a detailed experimental study of the three-jet
$b \bar b g$ final state could be useful for background calibration purposes. 
This is of particular interest for the kinematic configurations which
are enhanced, for example when an energetic gluon recoils against a quasi-collinear
$b \bar b$ pair. 
Recall also that the CDF collaboration is currently measuring exclusive
$b \bar b$ production at the Tevatron \cite{CDFb}.

\end{itemize} 

Note that in this paper we do not discuss the effects coming from collisions of {\it two} soft Pomerons, neither do we
address a possible contribution from central
inelastic production, see \cite{KMRProsp}. The reduction of such backgrounds
is controlled by imposing the missing mass equality, see \cite{DKMOR}. 
Note also that gluon radiation off the
screening gluon (labelled `$Q$' in Fig.~$\ref{fig:H}$) is numerically small \cite{myths}.

\subsection {Properties of the leading-order $gg^{PP}\ra\bb$ background process}

As we have discussed, an important 
advantage of the $p+(H\ra\bb)+p\,$ signal is that there exists
a $J_z=0$ selection rule, which requires the LO
$gg^{PP}\ra\bb$ background to vanish in the limit of
massless quarks and forward outgoing protons. 
However,
in practice, LO background contributions remain, see \cite{DKMOR}. The prolific
$gg^{PP}\ra gg$ subprocess can mimic $\bb$ production when the outgoing gluons are misidentified as $b$ and $\bar{b}$ jets.
Assuming the expected $1\%$ probability of misidentification, and
applying a $60^\circ<\theta<120^\circ$ jet cut, gives a
background-to-signal ratio $B/S \sim 0.2$ \cite{DKMOR, bh75}. 
(Here and in what follows,  we
assume for reference that the mass window over which we collect the signal is $\Delta M\sim 3\sigma=3$ GeV).
\footnote {Such a background is practically negligible in the case of the PC
since it must be mediated by the higher-order `box' diagrams.} 

Secondly, there is an
admixture of $|J_z|=2$ production, arising from non-forward going
protons, which gives $B/S \sim 0.05$, see Section~2.5.\footnote {Analogous to this in the PC case is the contribution from the initial photon state with  $|J_z|=2$, 
which may constitute a non-negligible source of background.}

Thirdly, in reality the quarks have non-zero mass and 
there is a contribution to the $J_z=0$ cross section of order
$m_b^2/E_T^2$, where $E_T$ is the
transverse energy of the $b$ and $\bar{b}$ jets. In \cite{DKMOR,bh75} 
the contribution from this source  was estimated 
as  $B/S \sim 0.2$. However the higher-order QCD effects may strongly
affect this result.
First, there is a  reduction coming from
the self-energy insertion into the $b$-quark propagator, that is 
from 
the running of the $b-$quark mass from $\overline{m}_b
(m_b)$  to its 
value $\overline{m}_b (M_H)< \overline{m}_b
(m_b)$ at the Higgs scale.  Here
$\overline{m}_b (\mu)$ 
is the running $b-$quark mass in the $\overline{\rm MS}$ scheme
\cite{BBDM}. It is known that in the $H\to b
\bar b$ decay width these single logarithmic (SL)
 ($ \alpha_S \ln \frac{M_H}{m_b}$) effects
diminish the corresponding Born result by a factor of approximately two
\cite{BL}.
Although this still requires a more formal proof,
we strongly believe that the same (factor of two) reduction
applies in the case of  mass-suppressed  $J_z=0$
binary reactions $gg^{PP}\to b \bar b$
and  $\gamma\gamma \to b \bar b$ at large angles.

There is another (potentially important) source of uncertainties 
in the evaluation of the rate of exclusive $b \bar b$
production at  $J_z=0$.
This is related to the so-called non-Sudakov form factor in the cross section
$F_q$ which arises from virtual diagrams with gluon exchange
between the final quarks (or initial gluons), see
\cite{fkm,JT1,melles}. In the $\gamma\gamma$
case the double logarithmic (DL) approximation to $F_q$ has the
form
\be
F_q (L_m) \; = \; \sum_n \: c_n \left ( \frac{\alpha_S}{\pi} \:
L_m^2 
\right )^n
\label{eq:p1}
\ee
with
\be
L_m \; \equiv \; \ln \left ( \frac{M_H}{m_b} \right ), \quad\quad 
\label{eq:p2}
\ee
 $c_0 = 1$ and $c_1 = -8$ \cite{JT1} so that the second
(negative) term in (\ref{eq:p1}) is anomalously
large and dominates over the Born term for $M_H \sim 100$~GeV.  This 
dominance undermines the results of any analysis
based on the one-loop approximation.
The physical origin of this non-Sudakov form factor was elucidated in
\cite{fkm} where its explicit calculation in the two-loop approximation
was performed. It was also shown that for reliable calculations of the DL effects
the two-loop calculation should be sufficient. This was subsequently confirmed by a
more comprehensive all-orders study \cite{melles}.
As is well known, there are other DL effects (the so-called Sudakov logarithms \cite{sud}) that arise from virtual soft gluon exchanges.
Their contribution depends on the particular kinematics in the final state.
As discussed in \cite{fkm}, in the case of quasi-two-jet configurations
Sudakov and non-Sudakov effects can be with  good accuracy
factorised, because they correspond to very different virtualities
of the internal quark and gluon lines.
For the final state radiation, the Sudakov effects (both for the signal and for the background) can be implemented in parton
shower Monte Carlo models in a standard way. For the $gg^{PP}$ initial state, the Sudakov factors are explicitly incorporated in the unintegrated gluon densities, see \cite{KMRProsp,KMR}.
Currently, for Higgs production at the PC the DL factors
are accounted for by the simulation programme used for generating background events,
see \cite{srj}.
Unfortunately, from a phenomenological perspective, it seems to be potentially 
dangerous to rely on the DL results, since  experience shows
that formally subleading SL corrections may be numerically important. We plan to address this issue
in future. 

Nevertheless, as a first step in understanding the situation in the $pp$ case 
it is instructive to evaluate the size of the DL effects
for the $gg^{PP}\to b \bar b g$ reaction.
Recall that in the photon-photon case the two-loop expression
for $F_q$ takes the form \cite{fkm} 
\be
\label{eq:p3}
F_q  =  (1 - 3 {\cal F})^2 \: + \: \frac{{\cal F}^2}{3} \: \left (1
\: + \: \frac{C_A}{C_F} \right ),  
\ee 
with\footnote{Note that an additional problem is that while in the Born cross section
it is natural to evaluate  $\alpha_S$ at the hard scale $M_H$,
 for ${\cal F}$ we have no
reason to adopt this prescription. The existing PC generators  
do not take into account possible differences in the scale of
$\alpha_S$ for the 
different quantities.}
\be
{\cal F} \; = \; \frac{\alpha_S}{\pi} \: C_F \: L_m^2 .
\label{eq:p4}
\ee
The corresponding one-loop result for the $gg^{PP}$ initiated process is
\be
\label{eq:p5}
F_g\; =\; 1\; -\; (2C_F\; +\; 4N_c){\cal F}/C_F\;\sim\; 
(1\; -\; (C_F\; +\; 2N_c){\cal F}/C_F)^2
\ee

Unfortunately, due to the large colour coefficients the one-loop
DL contribution becomes larger than the Born term, and the final result
will be strongly dependent on the NNLO  contribution as well as
on the scale $\mu$ at which the QCD coupling  $\alpha_S$
is evaluated and on
the running $b-$quark mass.  
It seems plausible to choose the scale $\mu \sim M_H/2$
(since we are interested in the $b\bar b$ background at $s=M^2_{b\bar b}=M^2_H$) 
for the Born amplitude, but a lower scale $\mu \sim \sqrt{M_H m_b}$ 
for the factor ${\cal F}$ in Eq.~(\ref{eq:p4}) which originates
in the region where the quark propagators are close to the mass-shell.
The NLL can be effectively incorporated in Eqs.~(\ref{eq:p3},\ref{eq:p5})
by introducing a scale factor $c$ in the argument of the
logarithm, that is by replacing the ratio ${M_H}/{m_b}$ by ${cM_H}/{m_b}$.  
It follows from the comparison with the complete one-loop calculation
\cite{JT1} for the process  $\gamma\gamma (J_z=0) \to b \bar b$,
that the scale factor $c\simeq0.5$.
This looks quite reasonable if we account for the kinematical configuration.

 Without the complete result for the higher-order 
radiative corrections corresponding to the 
$gg^{PP}\to b\bar b$ amplitude, it is impossible to make a firm prediction.
To gain an insight into the size of the possible effects, we make the assumption that the same scale factor $c=1/2$ 
is valid in this case as well. Then choosing $\alpha_S=\alpha_S(M_Hm_b)\sim 0.15$, the  
value of the correction $(1 + 2N_c/C_F){\cal F}\simeq 2.5$ exceeds the Born term.
In other 
words, the whole amplitude changes sign and the background cross section becomes
a few times 
larger than the Born expectation.

Accounting for the running $b$-quark mass, the expected non-Sudakov correction
factor can be approximated by
\begin{equation}
\label{eq:sud-g}
\left[1\ -\ \frac{m_b(M_Hm_b)}{m_b(M_H^2)}
(C_F+2N_c)\frac{\alpha_s(M_Hm_b)}{\pi}\ln^2\left(\frac{c\cdot
M_H}{m_b(M_Hm_b)}\right)\right]^2 \/ .
\end{equation}
For the  $\gamma\gamma$ case we have a similar expression
\begin{equation}
\label{eq:sud-gam}
\left[1\ -\ \frac{m_b(M_Hm_b)}{m_b(M_H^2)}
3C_F\frac{\alpha_s(M_Hm_b)}{\pi}\ln^2\left(\frac{c\cdot
M_H}{m_b(M_Hm_b)}\right)\right]^2
\end{equation}
If we take these formulae literally we would conclude that
in the $gg$ case the quasi-two jet cross section is a factor of 2 larger than the `naive' (but frequently used) Born prediction, calculated with $\alpha_S(M_H)$ and the $b-$quark pole mass. Similarly,  for the
  $\gamma\gamma$ process the result is about 5 times lower than such a naive Born estimate.

While for the $\gamma\gamma$ case the estimated effect seems to be
 reasonably justified,
for the gluon-initiated process it can serve only as a rough illustration of the possible size of the effect.
The actual results will depend
 crucially on 
the value of the scale factor $c$ (i.e. the NLL contributions) and on the
 specific choice of the arguments of the running coupling $\alpha_S$ and $b$-quark
mass. The main purpose of the exercise was to demonstrate that, currently (or, at least, before a complete 
one-loop result becomes available), the $gg^{PP}\to b\bar b$ cross section can be estimated to
no better than an order of magnitude accuracy.

\subsection{Quasi-two-jet-like radiative background events}

As was first found in \cite{JT1} for the $\gamma\gamma$ case,
there is an additional NNLO contribution
which is not mass-suppressed and is potentially important especially
for large energies. It comes from the QHC box diagrams. 
This piece cannot be evaluated in terms of the tree-level
amplitudes using the cutting rules
in 4 dimensions. A systematic method for calculating such amplitudes
 in the massless limit is based on  generalised
unitarity in $D$-dimensions, see for instance \cite{D}.
An explicit calculation in \cite{JT1} using dimensional
regularization of the NNLO process
 $\gamma\gamma \to q \bar q$ at large angle $\theta$
gives

\be
\label{eq:p6}
\frac{d\sigma^{NNLO}}{d\sigma_{\rm Born}}(\gaga\to q\bar q,J_z=0)\; =\;\frac{\alpha_S^2}
{32}C^2_F\frac s{m^2_b}\cos^2\theta(1-\cos^2\theta)
\ee
with
\be
\label{eq:p7}
\frac{d\sigma_{\rm Born}}{d\cos\theta}(\gaga\to q\bar q,J_z=0)\; =\; 
\frac{12\pi\alpha^2Q^4_q}s
\cdot\frac{\beta(1-\beta^4)}{(1-\beta^2\cos^2\theta)^2},
\ee
where $\beta \equiv \sqrt{1 - 4 m_q^2/s}$, and $m_q$ and $Q_q$
are the mass and 
electric charge of the quark respectively.\footnote {In \cite{fkm}
 there are some confusing statements
regarding the properties of the  $\gamma\gamma \to b \bar b$ amplitude in the complex plane
and the interpretation of the one-loop amplitude result
of \cite{JT1}. But these do not affect the actual formulae.}
Note that as is easily seen from  Eq.~(\ref{eq:p7}),
the NNLO elastic cross section vanishes at $\theta=\pi/2$.
This is a consequence of the rotational invariance
about the quark direction at $180^\circ$ and the identity of the photons (see Ref.~\cite{BKSO}) and remains valid  in the absence of radiation at all orders in 
$\alpha_S$.
This is also true for the $gg^{PP}\to q\bar q$ process.
The ratio in (\ref{eq:p7}) reaches its maximum at $\theta=\pi/4$ where
\begin{equation}\label{nnlo} \frac{d\sigma^{NNLO}}{d\sigma_{Born}}(\gamma\gamma\to q\bar q,J_z=0)\; 
=\;\frac{\alpha_s^2}
{72\pi^2}\frac s{m^2_b} \/ .
\end{equation}
Accounting for the running $b-$quark mass and the NLLO
effects discussed in the previous Section, we conclude that even at $M_H\simeq 140$~GeV these NNLO contribution to the cross section could not exceed 0.1 of the modified Born term. The NNLO elastic 
$c\bar c$ contribution is 16 times larger, and at $M_H\simeq 130$~GeV becomes comparable with the modified $b\bar b$ exclusive term. However with a reasonably good  experimental $c-$quark rejection this background can be strongly reduced without seriously degrading the $H\to b\bar b$ signal.

Using  the existing results for the one-loop amplitudes of the  $gg\ra\bb$ 
process in the massless limit (see for example \cite{adr,nigel})
we can write down the corresponding NNLO expression for the ratio of the
 $gg^{PP}\ra\bb$ subprocesses  as

\be
\label{eq:p8}
\frac{d\sigma^{NNLO}}{d\sigma_{\rm Born}}(gg^{PP}\to q\bar q,J_z=0)\; =\;\frac{(C_F-N_c)^2}{C_F^2}\cdot
\frac{d\sigma^{NNLO}}{d\sigma_{\rm Born}}(\gaga\to q\bar q,J_z=0) \/.
\ee
Note that the appearance of the $(C_F-N_c)^2$ factor in Eq.~(\ref{eq:p8}) is not accidental.
It is a consequence of  supersymmetry requiring the vanishing of such helicity
amplitudes in a supersymmetric
theory, which happens if we put the fermions in the adjoint
representation (gluinos).\footnote{We are grateful to Lance Dixon
for an illuminating discussion of the properties of helicity amplitudes
in a supersymmetric theory.}
This is in marked contrast with the combination
$(C_F+2N_c)^2$ that appears in Eq.~(\ref{eq:p5}), where the result is of a purely
classical nature and supersymmery arguments cannot be applied.
Note that in the massive quark case, even if we were  to consider altering its colour representation
from the fundamental to the adjoint representation, we cannot
put it into the same supersymmetric multiplet with the 
massless gluon \cite{ld}.

\subsection {Admixture of $|J_z|=2$ production, arising
from non-forward going protons} 
In the exact forward direction, the $J_z=0$
selection rule is just a consequence of the s-channel
helicity conservation for the forward protons.
For the non-zero
transverse momenta of the outgoing protons ($p_{1,t},p_{2,t}$)
some admixture of the $|J_z|=2$ component appears.
Its value is
controlled by the orbital momentum transfered through the Pomeron
(i.e. through the colour singlet $t$-channel two gluon exchange) that is by
the product $(p_{t,i}\cdot r_t)$, where $i=1,2$, and the effective
Pomeron size $r_t\simeq 1/Q_t$ is driven by the inverse transverse
momentum $Q_t$ in the gluon loop. Thus, this admixture of the $|J_z|=2$
states can be evaluated as \cite{KMRmm}
$2p_{1,t}p_{2,t}/Q^2_t$ in the amplitude or
\begin{equation}
\label{eq:j1}
\frac{(2p_{1,t}p_{2,t})^2}{Q^4_t}
\end{equation}
for the cross section.
An additional factor $2$ arises from the
azimuthal angular averaging ($Q_j Q_k \to
\delta^{(2)}_{j,k}Q^2_t/2$).

Technically the polarisation structure is as follows.
In the equivalent gluon approximation the polarisation vector
of the active gluon is proportional to its transverse momentum
$e_{\mu,i}\simeq (p_{t,i}-(-1)^i Q_t)_\mu/x_i$ ($i=1,2$).\footnote{
Alternatively, the same result can be obtained in the LO using the planar gauge
 $(n_\mu\cdot A^a_\mu)=0$ with the gauge 4-vector $n_\mu$ parallel to
the 4-momentum of the centrally produced system $M$.}
Thus the product of two polarisation vectors can be written as
$$e_{\mu,1}e_{\nu,2}\ =\
\frac{(p_{t,1}-Q_t)_\mu(p_{t,2}+Q_t)_\nu}{x_1x_2}  \/ ,$$
\begin{equation}
\label{eq:j2}
e_{\mu,1}e_{\nu,2}
\ \propto\
p_{t,1,\mu}p_{t,2,\nu}-Q_{t,\mu}Q_{t,\nu}+
(Q_{t,\mu}p_{t,2,\nu}-p_{t,1,\mu}Q_{t,\nu})  \/ .
\end{equation}
After the ($\vec{Q}_t$) angular integration the last term in
(\ref{eq:j2}) vanishes while the second term gives
$-\delta^{(2)}_{\mu\nu}Q^2_t/2$. In terms of helicity amplitudes,
$\delta^{(2)}_{\mu\nu}$ corresponds to a pure $J_z=0$ state. On the
other hand, the first term in (\ref{eq:j2}), after the averaging over
$p_{t,i}$ directions, generates the $J_z=0$ and $|J_z|=2$ states with
equal probabilities.
Unfortunately, the amplitude with an extra $1/Q^2_t$ factor becomes less
convergent at small $Q^2_t$. The dominant contribution comes from the
region of relatively
low $Q_t\sim $~GeV (and even lower for the Tevatron energies). Therefore,
we cannot guarantee the precision of the numerical evaluation. Using the
MRST99
partons \cite{MRS99} we expect the $|J_z|=2$ admixture for central exclusive production
of the state with the mass $M\sim 120\; - \; 160$~GeV 
to be about 5\% at Tevatron energies and $\sim 1.5$\%
at the LHC ($\sqrt s=14$~TeV) \cite{KMRmm}.  

There is good news however. It is worth noting that the $|J_z|=2$ 
contribution to the $gg^{PP}\to q \bar q$
background is additionally suppressed numerically (by a factor of, at least,  $\sim 0.2$).\footnote {This suppression was not accounted for in \cite{DKMOR,KMRmm}. This provides added value to the improvement of the signal-to-background situation in the $ b \bar b$-case. }.
In order to gain an insight into the origin of
this  additional suppression, we note that
the cross section vanishes at $\theta=\pi/2$
in the $q \bar q$ rest frame (neglecting the proton transverse momenta in comparison
with the transverse energy of the quarks). This follows from the identity
of the incoming gluons (protons) and invariance with respect
to the  $180^\circ$ rotation  about the quark direction.
This in  turn, causes the cross section to be proportional to $\cos^2\theta$. This phenomenon is also seen in the vanishing of the NNLO non-radiative amplitude at  $\theta=\pi/2$ considered in Section~2.3, and in the soft radiation off the screening gluon, considered in \cite{myths} (see also the discussion in \cite{BKSO}).

\subsection{NLO radiation close to the beam directions}

The NLO subprocess $gg^{PP}\to b \bar b g$ can also avoid the
$J_z=0$ selection rule.
Extra gluon radiation in the beam
direction goes into the beam pipe, and 
cannot be observed directly. Therefore, experimentally, the event
may look like central exclusive production. 
There are two main consequences of this extra gluon radiation:
a) the system  $M_8$ which is centrally produced via $gg\to M_8$
fusion is now in the colour octet state (which we label by the symbol `8'),
and (b) the $J_z=0$ selection rule, which suppresses
the $b\bar b$ LO QCD production in the genuine central exclusive event,
becomes redundant.

 Let us discuss this point in more detail.
The emission of a low $q_t$ extra gluon is strongly suppressed
due to the distructive interference between the amplitudes where the gluon $q_\mu$
is emitted from the right (active) or from the left (screening)
$t$-channel gluons in Fig.~$\ref{fig:H}$.
Therefore, it is sufficient to consider the case when
$M_H \gg q_t \gg Q_t$, and the polarisation of the active gluon
(participating in the $gg\to M_8$ fusion process) is directed along the new vector
$\vec{q}_t$.
Assuming that the extra gluon with momentum $q_\mu$ is
emitted from
the lower active gluon ($i=2$), the
polarisation structure becomes
\begin{equation}
\label{eq:j3}
e_{\mu,1}e_{\nu,2} \ \propto\
(p_{t,1,\mu}-Q_{t,\mu})(q_{t,\nu}+Q_{t,\nu})\simeq
p_{t,1,\mu}q_{t,\nu}-Q_{t,\mu}q_{t,\nu}-\delta^{(2)}_{\mu\nu}Q^2_t/2\ ,
\end{equation}
where the last term corresponds to the $J_z=0$ state and the first
term corresponds to the hard subprocess, which
after the $\vec q_t$ and $\vec p_{t,1}$ averaging
looks like the usual fusion of two unpolarised gluons (with equal
probabilities for the $J_z=0$ and $|J_z|=2$ initial states).
The contribution generated by the second term is more complicated.
At first sight it should vanish after the integration over the azimuthal
$\vec Q_t$ angle, however due to the factor $(\vec Q_t+\vec
q_t)^2$ in the denominator of the amplitude some component of the
momentum $Q_{t,\mu}$ in the $q_{t,\mu}$ direction still survives.
In the limit of $q_t \gg Q_t$ this
leads to a contribution $\sim q_{t,\mu}q_{t,\nu}Q^2_t/q^2_t$
to the right-hand side of Eq.~(\ref{eq:j3}), which again contains the $J_z=0$ and
$|J_z|=2$ states with equal probabilities.

Thus, the ratio of the $|J_z|=2$ to $J_z=0$ contributions
to the cross section for this process may be evaluated as
\begin{equation}
\label{eq:j4}
  \frac{\sigma(|J_z|=2)}{\sigma(J_z=0)}\ =\
\frac{<Q^2_t>^2\ +\ <p^2_t><q^2_t>}
{2<Q^2_t>^2\ +\ <p^2_t><q^2_t>} \/ .
\end{equation}
We might expect that the probability to emit such an extra gluon
would contain a double logarithm,
but this does not happen. First, as will be discussed in Section~\ref{sec:numer}, for the massless $b-$quark case the
soft gluon emission $gg^{PP}\to b \bar b g$ is suppressed by a factor $(E_g/M_{b \bar b})^4$ and therefore
gives no logarithm. On the other hand, the collinear
  logarithm is limited by the angular acceptance of the detector.
Any gluon with a sufficiently large $q_t$ will be observed in the Central
  Detector. 
Moreover, up to pseudorapidities $\vert \eta\vert \sim 6 - 7$ the extra gluon jet will be observed experimentally
in a Forward Detector. Such events with a third jet will be easily distinguished from the $H\to b \bar b$ decay.
Next, if the energy of the gluon $q$  exceeds the
mass resolution then there will be no matching between the
  missing mass calculated from the momenta of the outgoing forward
  protons and the mass $M_8$ measured in the Central Detector.
Assuming the mass resolution $\Delta M_{b \bar b}\sim 20$~GeV, we require that for central $b \bar b$ production the energy of the third (forward) gluon jet must be less than 40~GeV, and to get $\vert\eta\vert > 6$ such a jet must have a very small transverse momentun, $q_t < 2 E_g \exp(-6) = 0.2$~GeV. The production of such a low $q_t (\ll Q_t)$ gluon is strongly suppressed by the interference between the emissions of the active gluon $q_t$ and the screening gluon $Q$ (see Fig.~\ref{fig:H}). This contribution becomes smaller than the admixture of the $\vert J_z \vert = 2$ states.

\def\gaga{\gamma\gamma}

\section{The $gg,\gaga \to q\bar q g $  $|J_z|=0$ colour singlet hard process}.
\label{sec:calc}

In this section we will present results for the matrix elements squared for the colour singlet hard 
scattering processes $gg\to q\bar q g $ and $ \gaga\to q\bar q g $. Since we will be using these results in situations where the momentum transferred in the hard scattering is much larger than the 
$b-$quark mass, we will set $m_q = 0$.  We will compare our results with the corresponding full spin- and colour-summed amplitudes in order to exhibit the different limiting behaviours. 

In fact the spin- and colour-summed matrix element squared for the $2\to 3$ process $g(p_1) + g(p_2)  \to 
g (p_3) q(p_4) \bar q (p_5)$ process has been known for a long time \cite{Berends:1981rb} and has a relatively simple analytic form:
\bea
\sum\vert{\cal M}\vert^2 & = & \frac{g_s^6}{4N^2(N^2-1)}\; \left( \frac{a_1b_1(a_1^2+b_1^2) + a_2b_2(a_2^2+b_2^2) + a_3b_3(a_3^2+b_3^2)}{a_1a_2a_3b_1b_2b_3}\right)  \nonumber \\
&& \left[ \frac{s}{2} + N^2 \left( \frac{s}{2} - \frac{a_1b_2+a_2b_1}{d_{12}}-\frac{a_2b_3+a_3b_2}{d_{23}}
-\frac{a_3b_1+a_1b_3}{d_{13}}\right)  \right. \nonumber \\
&& \left. + \frac{2N^4}{s} \left( \frac{a_3b_3(a_1b_2+a_2b_1)}{d_{23}\; d_{13}}+\frac{a_1b_1(a_2b_3+a_3b_2)}{d_{12}\;d_{13}}
       +\frac{a_2b_2(a_3b_1+a_1b_3)}{d_{12}\;d_{23}} \right)  \right] 
\label{eq:gggbbanal}
\eea 
where $a_i = p_i \cdot p_4$, $b_i = p_i \cdot p_5$, $d_{ij} = p_i \cdot p_j$ ($i,j = 1,...3$) and $s = 2 p_4 \cdot p_5$. 
An averaging over initial spins ($1/4$) and colours ($1/(N^2-1)^2 = 1/64$) has been performed.
In fact the above spin-summed amplitude squared comprises 12 distinct non-zero helicity combinations, 4 of which 
\begin{equation}
(++;--+), \ 
(++;-+-), \ 
(--;+-+), \ 
(--;++-)
\end{equation}
(in an obvious notation) correspond to a $J_z = 0$ initial state, while the remaining 8
\bea
&(-+;+-+), \ 
(-+;--+), \ 
(-+;++-), \ 
(-+;-+-), & \nonumber  \\
& (+-;+-+), \ 
(+-;--+), \ 
(+-;++-), \ 
(+-;-+-) &  
\eea
correspond to a $\vert J_z\vert  = 2$ initial state. Note that in all cases 
$\lambda_q = - \lambda_{\bar q}$, 
corresponding to helicity conservation along the fermion line. The other important point to note is that all the above combinations are MHV amplitudes \cite{mhv1}, in the sense that the sum of the (five) helicities is always $\pm 1$. At this order there are no NMHV amplitudes or higher. MHV $gggq\bar q$ scattering amplitudes have a very simple analytic form, see the Appendix, and so with appropriate colour weightings and momentum permutations, the colour singlet $J_z = 0, \pm 2$ matrix elements squared can be easily constructed.  The result is 
\bea
\sum\vert{\cal M}\vert^2(J_z=0; \mbox{colour singlet}) & = & \frac{2}{9}\; \sum_{h=1}^4 \; \Big\vert z(1,2,3,h)+z(2,1,3,h)+z(3,2,1,h)
\nonumber \\
& & +z(3,1,2,h)         - \frac{1}{8}\left( z(1,3,2,h)+z(2,3,1,h) \right) \Big\vert^2
\eea
where spin and colour averaging factors are included. The $ z$ factors are given in the Appendix.
Expressions for the other spin and colour combinations can also be written down in terms of the $z(i,j,k,h)$ factors. However these are more lengthy and so will not be presented here.

Compact analytic expressions exist for the corresponding $\gaga \to g q \bar q$  spin summed and $J_z = 0$ amplitudes squared. In the notation of Eq.~(\ref{eq:gggbbanal}), 
\bea
\sum\vert{\cal M}\vert^2_{\gaga}(\mbox{spin summed}) & = & 8 g_s^2 e^4\; \frac{s}{2} \; \left( \frac{a_1b_1(a_1^2+b_1^2) + a_2b_2(a_2^2+b_2^2) + a_3b_3(a_3^2+b_3^2)}{a_1a_2a_3b_1b_2b_3}\right) \/ , \nonumber \\
\sum\vert{\cal M}\vert^2_{\gaga}(J_z=0)  & = & 8 g_s^2 e^4\; \frac{s}{2} \; \left( \frac{a_3^2+b_3^2}{a_1a_2b_1b_2}\right) \/ .
\label{eq:gagagbbanal}
\eea 

\section{Numerical results and discussion}
\label{sec:numer}

In this section we consider some of the properties of the amplitudes presented in the previous section and, in particular, focus on the differences between the spin-summed (unpolarised) and $J_z=0$ cases. The differences are most dramatic for the kinematic configuration in which the final-state gluon is {\it soft}. 
It follows from the $E_g\to 0$ limit of Eq.~(\ref{eq:gagagbbanal}) that
the matrix element squared for the $J_z=0$ case is proportional to $E^2_g$,
while in the unpolarised case it exhibits the standard  $1/E^2_g$ behaviour.
This is because the first two terms in the bracket on the right hand side
in the unpolarised case, which are responsible for the leading infrared behaviour in the soft-gluon limit, are absent in the $J_z=0$ case. The net difference is four powers of $a_3$ or $b_3$, equivalently $E_g^4$. Numerical calculation shows that exactly the same behaviour is found in the $gg$ scattering cases.

In terms of the cross sections in the soft gluon limit,
\begin{equation}\label{C4} 
\frac{d\sigma (J_z=0)}{dE_g}\sim E^3_g\/, 
\end{equation} 
while in the unpolarised case we arrive at the the standard infrared behaviour 
\begin{equation}\label{C5} 
\frac{d\sigma_{{\rm unpol}}}{dE_g}\sim \frac{1}{E_g}\/.
\end{equation} 
Such behaviour is rooted in the Low-Burnett-Kroll (LBK) \cite{LBK} theorem
(see also \cite{BKSO, DKS, kmrRj}).
According to the LBK theorem, for radiation of a soft gluon
with energy fraction $x_g \ll 1$, the radiative matrix element $M_{\rm rad}$
may be expanded in powers of the scaled gluon energy $x_g={E_g}/{E_b}$
\begin{equation}\label{lbk} 
 M_{\rm rad} \sim  \frac{1}{x_g} \sum_{n=0}^\infty C_n x_g^n,
\end{equation}
where the first two  terms, with coefficients $C_0$ and $C_1$ (which correspond
to long-distance radiation), can be written
in terms of the non-radiative matrix element $M_B$.
 The application of these classical results is especially transparent when 
the cross sections are integrated over the azimuthal angles. 
Then the non-radiative
process depends only on simple variables, such as the centre-of-mass energy.
When $M_B =0$, the expansion
starts from the non-universal $C_2 x_g^2$ term, which corresponds to non-classical
(short-distance) effects, not
related to $M_B$.
This is exactly the case
for the  $J_z=0$ Born amplitudes
which vanish for massless
quarks for both $\gamma\gamma$ and $gg^{PP}$ processes.\footnote{Actually, 
the $J_z=0$ non-radiative matrix element vanishes for {\it any} colour state of 
two gluons since it corresponds to a Maximally Helicity Violating (MHV)
situation, see for example \cite{MP}.}
On the other hand, in the soft limit the unpolarised
result is dominated by the  non-vanishing
non-radiative amplitudes, either $(-+;+-)$, $(-+;-+)$ 
or $(+-;+-)$, $(+-;-+)$, and in this case the matrix element squared,
$|M_{\rm rad}|^2$, is proportional to $1 / E^2_g$. Recall, however,
that the mass-suppressed contributions to the $J_z=0$ amplitudes
will induce the normal infrared behaviour, see for example \cite{BKSO}, i.e.
\begin{equation}\label{C4m}
\frac{d\sigma (J_z=0)}{dE_g}\sim \frac{m_b^2}{s}\frac{1}{E_g}\/.
\end{equation}

\begin{figure}
\begin{center}
\centerline{\epsfxsize=0.8\textwidth\epsfbox{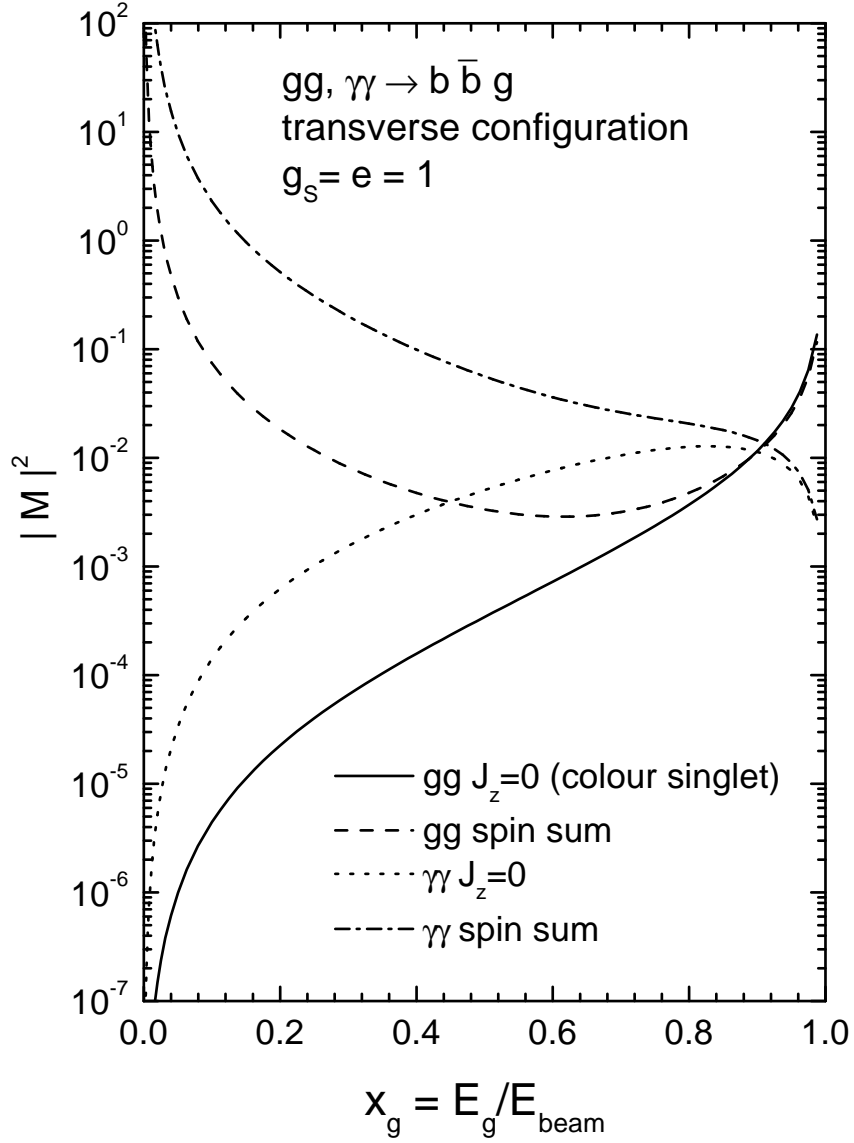}}
\vspace{-2cm}
\caption{Dependence of the $gg, \gaga \to  b {\bar b} g$ matrix element squared on the final-state gluon fractional energy, when all three particles are produced in the transverse plane and the $b$ and $\bar b$ have equal energy. Note that the $b-$quark mass is set to zero.}  
\label{fig:gggbb3}
\end{center}
\end{figure}

The above behaviour is illustrated in Fig.~\ref{fig:gggbb3} where we plot\footnote{For display purposes, the QCD and QED couplings $g_s$ and $e$ are set to 1 in this plot.}  the matrix elements squared as a function of the scaled gluon energy $x_g = E_g / E_{\rm beam}$. For simplicity, we work in the centre-of-mass frame with all final state particles in the transverse plane and equal energies for the (massless) quark and antiquark. In this configuration, $x_g \to 0$ is the soft gluon limit, $x_g = 2/3$ is the (transverse) `Mercedes' configuration, and $x_g \to 1$ corresponds to a transverse gluon balanced by a collinear $q \bar q$ pair. 

The expected LBK behaviour discussed above ($x_g^2$ versus $x_g^{-2}$) is clearly seen in the $x_g \to 0$ limit. Note also that  the $gg$ (but not the $\gaga$) amplitudes become singular as $x_g$ approaches the kinematic limit at $x_g = 1$. This is the collinear singularity caused by the production of two final-state back-to-back gluons, one of which splits into a quark-antiquark pair, i.e. $g \to q \bar q$. For these kinematics, this is manifest as a $(1-x_g)^{-1}$ singularity in the matrix elements squared.   There is no analogue in the $\gaga$ case, and indeed here the amplitudes vanish in the $x_g \to 1 $ limit. since at $x_g \to 1$ the quark and antiquark go in exactly the same direction and their electric charges screen each other. Thus the coupling of the photon to the $q \bar q$ pair vanishes. 

More generally, we can study the behaviour of the matrix elements squared when all three particles lie in the transverse plane by using the Dalitz-plot variables $x_q$ and $x_{\bar q}$, the scaled energies of the final state quark and antiquark respectively, with $x_q +  x_{\bar q} + x_g = 2$. For the $J_z = 0$ $\gaga\to q\bar q g$ case, it is straightforward to show that
\begin{equation}
\sum\vert{\cal M}\vert^2_{\gaga}(J_z=0)  =  \frac{256 g_s^2 e^4}{s}\; \left[ (1-x_q)^2 + (1-x_\qbar)^2 \right]  
\frac{x_q + x_{\qbar} - 1}{x_q^2 x_\qbar^2} \/. 
\label{eq:gagagbbfractions}
\end{equation} 
In the limit where $x_\qbar$ (or $x_q$) $\to 1$, an antiquark balances a collinear quark--gluon pair, and the matrix element squared becomes   
\begin{equation}
\sum\vert{\cal M}\vert^2_{\gaga}(J_z=0)   \to  \frac{256 g_s^2 e^4}{s}\; (1-x)^2 / x \/, 
\label{eq:gagagbbfractionslimit}
\end{equation} 
with $x = 1-x_g $ the fractional quark momentum of the quark--gluon pair. Note also that the matrix element squared vanishes when the quark and antiquark are emitted in the same direction ($x_q + x_{\qbar} = 1$), as anticipated above.

For the $gg$ scattering case, there is no simple analogue of (\ref{eq:gagagbbfractions}) for the whole Dalitz plot. However, the behaviour along the boundaries can be extracted and studied. We find  
\begin{equation}
\sum\vert{\cal M}\vert^2_{gg}(J_z=0)   =  \frac{ g_s^6}{s}\; \left[ (1-x_q)^2 + (1-x_\qbar)^2 \right]  
\frac{F(x_q,x_\qbar)}{x_q + x_{\qbar} - 1} \/, 
\label{eq:gggbbfractions}
\end{equation} 
where $F(x,y)$ is a non-singular function with constant values along the three sides of the Dalitz plot:
\bea
F  & = & \frac{64}{9}, \qquad \mbox{when}\ x=1\ \mbox{or}\ y=1 \nonumber \\
F  & = & 36, \qquad \mbox{when}\ x+y=1 \/ .
\label{eq:gggbbF}
\eea

Note the jump of the function $F$ at $x=1, y\to 0$ and $y=1, x\to 0$. These are the points where the soft antiquark (or quark) changes its direction, leading to a discontinuity in the classical coloured current.

To summarise, in the $gg$ $J_z = 0$ colour-singlet case, the only final-state\footnote{We note that we are considering here central production of all three final-state particles. There are of course additional singularities when the final-state gluon is emitted collinear with the incoming gluons, but these initial-state singularities are well understood.} singularity is when the gluon is emitted opposite a collinear quark-antiquark pair. In practical terms, this corresponds to the case when both $b$-quarks are contained within the same jet. Strictly speaking, this does not constitute a background to Higgs production, since the latter gives rise to two distinct $b$-jets. 

In contrast, the full spin-, colour-summed $gg$ amplitude has additional singularities when $x_q = 1$ or $x_\qbar = 1$. In fact the full singularity structure is exhibited in the empirical form
\begin{equation}
\sum\vert{\cal M}\vert^2_{gg}(\mbox{spin, colour summed})   =  \frac{ g_s^6}{s}\; \frac{G(x,y)}{(1-x_q)(1-x_\qbar)(x_q + x_{\qbar} - 1)} \/, 
\label{eq:gggbballfractions}
\end{equation} 
where $G$ is non-singular throughout the Dalitz plot. Note that when $x_q = x_\qbar = 1-x_g/2$ --- the kinematics of 
Fig.~\ref{fig:gggbb3} --- this reduces to 
\begin{equation}
\sum\vert{\cal M}\vert^2_{gg}(\mbox{spin, colour summed})  =  \frac{ 4 g_s^6}{s}\; \frac{G(1-x_g/2, 1-x_g/2)}{x_g^2 (1-x_g)} \/ . 
\label{eq:gggbballfractionssim}
\end{equation} 

What does this mean for jet cross sections? Recall that in $e^+e^- \to q \bar q g$ the `three-jet cross section' is defined by integrating the matrix element squared over a region away from the boundaries of the Dalitz plot, the exact region depending on the jet algorithm definition. For example, the $\ycut$ (JADE) algorithm defines the three-jet region by
\begin{equation}
1-x_q, 1-x_\qbar, x_q + x_\qbar + 1 \ > \ \ycut \/.
\end{equation}
Given the absence of collinear singularities in the $J_z = 0$, colour-singlet case when the gluon is emitted parallel to the quark or antiquark, we may therefore expect relatively more radiative three-jet events than in the $|J_z| = 2$ or spin-summed cases. Note, however, that the three jet cross section is still (logarithmically) singular in the $\ycut \to 0$ limit corresponding to the configuration $x_q + x_{\qbar} \to 1$  in which the final-state $b$ and $\bar b$ are collinear, i.e. $g + g \to g + g^* (\to b\bar b)$. However,  such radiatively generated quasi-two-jet events can be suppressed by requiring two distinct, spatially-separated $b-$tagged jets. Indeed applying a minimum-angle cut between the $b$ and $\bar b$ jets leads to a finite ($J_z = 0$, colour-singlet) three-jet cross section.

Note that in order to compare the matrix elements for the $\gamma\gamma \to b \bar b g$ and $gg \to b \bar b g$ reactions in Fig.~2 we set the couplings $g_s = e = 1$. Strictly speaking, at the leading order (tree level) at which we work, it is not known at what scale the couplings must be evaluated. However in actual cross section calculations it would appear natural to take two of the vertices at the large scale $\mu_{1,2}^2 \sim s/2$ and the third (outgoing gluon emission) vertex at a lower scale $\mu_3^2 \sim k_\perp^2$, i.e. $\vert M\vert^2 \propto \alpha_s^2 (s/2) \alpha_s(k_\perp^2)$, where $k_\perp$ is the transverse momentum of the final gluon with respect to the nearest (quark, antiquark or beam) jet direction.

In summary, provided that two distinct $b-$jets are required, then the dominant background arising from $gg \to b \bar b g$ production in the $J_z = 0$, colour-singlet case corresponds to three-jet production. The same is true for $\gaga$ production.
 Although we have concentrated our analysis on production in the transverse plane, this conclusion is still valid for central production, i.e. where the final state jets are restricted to a central region in rapidity.

\section{Summary}

 In the previous sections we have shown(see also
\cite {BKSO}) that, as a consequence of the Low-Burnett-Kroll theorem 
\cite{LBK}, when neglecting the quark mass, the differential distribution 
over the gluon energy in the $\gamma\gamma, gg^{PP}\to q\bar q g$ 
cross section is
\begin{equation}\label{C4a} 
\frac{d\sigma (J_z=0)}{dE_g}\sim E^3_g\/, 
\end{equation} 
in marked contrast to the Higgs or unpolarised case, where the 
cross sections exhibit the standard infrared behaviour 
\begin{equation}\label{C5a} 
\frac{d\sigma_{{\rm unpol}}}{dE_g}\sim \frac{1}{E_g}\/.
\end{equation} 
As a result,
the relative probability of the Mercedes-like configuration in the final 
$q\bar q g$ state for the $J_z=0$ background processes becomes unusually large.
We have derived explicit analytic expressions for the $\gamma\gamma, gg^{PP}\to q\bar q g$ amplitudes using MHV techniques, and calcualted some simple energy distributions to illustrate their generic behaviour. These amplitudes can easily be incorporated into more sophisticated Monte Carlo programmes to investigate background event rates in the presence of realistic experimental  cuts.

Finally, the approach of Ref.~\cite{DFKO} enables us to evaluate the difference 
between the charged multiplicities of the signal $N_S$ and 
Mercedes-like background events $N_{BG}^{\rm Merc}$ containing 
$b$-quarks for the $J_z=0$ initial state 
for both processes $\gamma\gamma$ and $gg^{PP}$ induced processes.
As is shown in \cite{DFKO}, 
\be 
\label{eq:C6} 
\Delta N = N^{\rm Merc}_{BG} (M_H)-N_S (M_H)=\\ 
       = N_{q\bar q}\left(\frac{M_H}{\sqrt{3}}\right)+\frac{1}{2} N_{gg} 
           \left(\frac{M_H}{\sqrt{3}}\right) - N_{q\bar q}(M_H) \/ .
\ee 
For example, for a 100~GeV Higgs boson, $2E_q^*=\frac{M_H}{\sqrt{3}}\simeq 
58~{\GeV}$, which corresponds to the energies of the existing measurements 
by TOPAZ and VENUS, see \cite{DFKO}. Substituting into Eq.~(\ref{eq:C6}) the 
corresponding experimental results for $N_{q\bar q}$ and the fits to 
the $gg$ multiplicity from \cite{DFKO}, we arrive at the 
multiplicity difference between the Mercedes-like background 
events and the $b\bar b$ signal, 
\begin{equation} 
\Delta N = 6.8 \pm 1.5 \; . \label{C7} 
\end{equation} 
A similar result ($\Delta N\sim 8.0$)  
appears if we use the existing $(udscb)$ direct data on the 
total charged multiplicity at the $Z^0$ pole and the corresponding 
number for the multiplicity of 3-jet events, see \cite{DFKO}.
Note that the multiplicity difference rises as $M_H$ 
increases.  
We expect that such a large effect could 
help to discriminate between the Higgs signal and background events 
containing $b$-quarks and in the analysis of the $b\bar b$ diffractive
events at the Tevatron.

\section*{Acknowledgements}
We thank Brian Cox, Albert De Roeck, Lance Dixon, Nigel Glover, 
Monika Grothe,
Victor Fadin, Jeff Forshaw, Valya Khoze,
Alan Martin, Klaus Moenig, Risto Orava,
Andrei Shuvaev and Georg Weiglein
for useful discussions. 
 MGR thanks the IPPP at the University of
Durham for hospitality. This work was supported by
the UK Particle Physics and Astronomy Research Council, by a Royal Society special
project grant with the FSU, by an INTAS grant 05-103-7515, by grant RFBR 04-02-16073
and by the Federal Programme of the Russian Ministry of Industry, Science and Technology
SS-1124.2003.2.

\newpage

\section*{Appendix: Helicity amplitudes for $gg \to gq\bar q$}

Here we outline the formalism used to calculate the $gg\to gq\bar q$ scattering amplitude 
discussed in Section~\ref{sec:calc}.
We denote the colour indices of the incoming gluons by $a,b$, and of 
the outgoing  gluon by $c$. The quarks colour indices are $j,k$.

The $gg \to gq\bar q$ matrix element, which depends on the helicities, 
$h_i$, and the 4-momenta, $p_i$, of the gluons and quarks, is given by the so-called 
dual expansion (see \cite{MP} and references therein)

 \begin{equation}
\label{mhv1}
M^{h_i}(p_i)_{jk}=\sum  (\lambda_a\lambda_b
\lambda_c)_{jk}~ z(a,b,c)\ ,
\end{equation}
where the sum is over the non-cyclic permutations of $a,b,c$.
The first factor has the same structure as if all the gluons were emitted from the quark line. The 
 $\lambda_i$ are the standard matrices of the fundamental representation of SU(3),
and are normalised as follows
\be
 {\rm Tr}( \lambda^a  \lambda^b)=  \frac{1}{2} \delta^{ab},
\ee
\be
 [\lambda^a,  \lambda^b]= i f_{abc} \lambda^c.
\ee
The colour-ordered subamplitudes, $z(a,b,c,)$, are only functions of the kinematical 
variables of the process,
i.e. the momenta and the helicities of the gluons. They
may be written in terms of the products of the Dirac bispinors, 
that is in terms of the angular (and square) brackets
\be
\langle ab \rangle~=~\langle p_a^-|p_b^+\rangle~=~\sqrt{|2p_ap_b|}e^{i\phi_{ab}},
\ee
\be
[ab]~=~\langle p_a^+|p_b^-\rangle~=~\sqrt{|2p_ap_b|}e^{i\bar\phi_{ab}},
\ee
where $2p_ap_b=s_{ab}$
is the square of the energy of the corresponding pair.  If both 4-momenta 
have positive energy, the phase $\phi_{ab}$ is given by
\be
\cos \phi_{ab}=\frac{p_a^xp_b^+-p_b^xp_a^+}{\sqrt{p^+_ap^+_bs_{ab}}},~~~~~~~
\sin \phi_{ab}=\frac{p_a^yp_b^+-p_b^yp_a^+}{\sqrt{p^+_ap^+_bs_{ab}}},
\ee
 with $p^+_i=p_i^0+p_i^z$, while
 the phase $\bar\phi_{ab}$ can be calculated using the identity $s_{ab}=\langle ab \rangle [ab]$.

   Finally, the only non-zero $J_z=0$ subamplitudes are
  \begin{equation}
\label{mhv2}
z(a,b,c;h)=ig_s^3 \frac{\langle qc\rangle \langle \bar qc\rangle 
\langle Ic\rangle ^2}{\langle \bar qq\rangle 
\langle qa\rangle\langle ab\rangle \langle bc\rangle 
\langle c\bar q\rangle } \/ .
\end{equation}
 
Here $g_s$ is the QCD coupling ($\alpha_s=g_s^2/4\pi$) and $I$ denotes the quark 
(or antiquark) which has the same 
helicity as the outgoing gluon $c$.  In particular, when $\lambda_a=\lambda_b=1$ 
while $\lambda_c=\lambda_q$  ($\lambda_{\bar q}=-\lambda_q$) the numerator takes the form
$\langle qc \rangle^3\langle \bar q c\rangle$. The expression (\ref{mhv2}) is written 
for the case of the incoming gluons with positive helicities. 
If we change the sign of all helicities, then we have simultaneously to replace
the $\langle ij \rangle$ brackets by the $[ij]$ brackets.

Note that in the formalism leading to (\ref{mhv2}) all the gluons are
considered as incoming particles; that is, the energies of gluon $c$ and both quarks are negative. In the case when one or two momenta in the product
$\langle ab \rangle$ have negative energy, the phase $\phi_{ab}$ is
calculated with minus the momenta with negative energy, and then
$n\pi/2$ is added to $\phi_{ab}$ where $n$ is the number of negative
momenta in the spinor product.

It is clear from (\ref{mhv2}) that in the limit of a soft outgoing gluon $c$
the cross section obtains an extra factor $E_g^4$ 
%(see (22,23)) 
since the numerator of the amplitude 
$$\langle qc \rangle \langle \bar q c\rangle\langle Ic\rangle ^2\propto E_g^2\/ .$$
This factor kills the soft gluon logarithm.
The collinear logarithm corresponding to the kinematics where gluon $c$ goes in,   say, 
the quark $q$ direction  may come only from the subamplitudes $z(c,a,b;h)$ and $z(c,b,a;h)$ 
where the
factor $\langle qc \rangle$  in the denominator of (\ref{mhv2}) provides the collinear 
singularity. However this singularity is cancelled by the analogous term in the numerator.
Thus the most dangerous background configuration (where the gluon is very hard to
separate from the quark jet) is not enhanced by any large logarithm.

It is straightforward to derive the analytical expression for the experimentally important 
 kinematic configuration where the gluon and quark directions are aligned. This involved keeping just the two subamplitudes, $z(c,a,b;h)$ 
and $z(c,b,a;h)$, with the quark helicity opposite to the gluon $c$ helicity 
(in order to have a larger numerator in (\ref{mhv2})).
This gives

\be
\sum |{\cal M}|^2(J_z=0; {\rm colour\  singlet})\ =\ \frac{4g_s^6}9 \frac{x_g^2s}{E^4_T}\/ .
\ee

\end{document}